\documentclass[12pt,preprint]{aastex}

\slugcomment{Submitted: October 2008; Revised: Nov. 25, 2008}

\shorttitle{Cepheid Distance to NGC~6822}
\shortauthors{Madore {\it et al.}}

\begin{document}


\title{The Cepheid Period-Luminosity Relation (The Leavitt Law) 
at Mid-Infrared Wavelengths: III. Cepheids in NGC~6822}


\author{\bf Barry F. Madore, Jane Rigby, Wendy L. Freedman \\ S. E. Persson,
Laura Sturch \& Violet Mager} \affil{Observatories of the Carnegie
Institution of Washington \\ 813 Santa Barbara St., Pasadena, CA
~~91101} \email{barry@ociw.edu, jrigby@ociw.edu, wendy@ociw.edu, 
persson@ociw.edu, lsturch@ociw.edu, vmager@ociw.edu}



\begin{abstract}
We present the first application of mid-infrared Period-Luminosity
relations to the determination of a Cepheid distance beyond the
Magellanic Clouds.  Using archival IRAC imaging data on NGC~6822 from
Spitzer we were able to measure single-epoch magnitudes for sixteen
long-period (10 to 100-day) Cepheids at 3.6$\mu$m, fourteen at
4.5$\mu$m, ten at 5.8$\mu$m and four at 8.0$\mu$m. The measured slopes
and the observed scatter both conform to the relations previously
measured for the Large Magellanic Cloud Cepheids, and fitting to those
relations gives apparent distance moduli of $\mu_{3.6}$ =
23.57$\pm$0.06, $\mu_{4.5}$ = 23.55$\pm$0.07, $\mu_{5.8}$ =
23.60$\pm$0.09 and $\mu_{8.0}$ = 23.51$\pm$0.08~mag. A
multi-wavelength fit to the new IRAC moduli, and previously published
BVRIJHK moduli, allows for a final correction for interstellar
reddening and gives a true distance modulus of 23.49 $\pm$ 0.03~mag
with E(B-V) = 0.26~mag, corresponding to a metric distance of
500$\pm$8~kpc.

\end{abstract}

\subjectheadings{Stars: Variables: Cepheids,  Galaxies: Individual (NGC 6822), Infrared: Stars,  Galaxies: Distances and Redshifts}
\vfill\eject
\section{Introduction}

We have embarked upon a program to recalibrate extragalactic distance
scale using Cepheid variables observed at mid-infafred
wavelengths. Two calibration papers (Freedman et al.  2008 and Madore
et al. 2009) gave the first-epoch and then the dual-epoch calibrations
of the four mid-infared period-luminosity relations at 3.6, 4.5, 5.8
and 8.0$\mu$m based upon 76 Cepheids in the Large Magellanic Cloud as
originally measured by the SAGE project (Meixner et al.
2006). Single-phase observations at these wavelengths have scatter
contributed in almost equal measure by the natural width of the
Cepheid instability strip and by the randomly sampled amplitudes of
the individual Cepheids themselves.  In the calibration papers we
found that  single-phase scatter of $\pm$0.14~mag is indicative of all
four of the mid-infrared PL relations. In addition to the small
scatter, the main advantage of observing Cepheids at these long
wavelengths is the overall insensitivity of these magnitudes to
interstellar extinction. For instance, compared to the optical B band,
the extinction suffered by stars measured at 4.5$\mu$m (say) is reduced
by a factor of 60 according to the extinction curve of Rieke \&
Lebovsky (1996, their Table 3). Thus any star with a line-of-sight
B-band extinction 0.60~mag or less will have mid-IR extinction
corrections of less than 1\%. For the majority of extragalactic
Cepheids reddenings are not all that high and this effectively means
that mid-IR distance moduli are effectively true distance moduli.

Hubble (1925) was the first discover Cepheids in NGC~6822. In Hubble's
own words NGC~6822 is ``the first object definitely assigned to a
region outside the galactic system'' based on ``familiar relations
such as those connecting periods and luminosities of Cepheids
...''. Hubble listed 11 Cepheids with periods ranging from 12 to 64
days.  Nearly half a century later, Kayser's (1967) comprehensive
examination of the resolved stars in NGC~6822 brought the number of
Cepheids up to 13 and modernized the study of this galaxy by
introducing both B and V observations of the main body of the
galaxy. Kayser also selectively used UBV data of bright field stars
along the line of sight to NGC~6822 to show that there was appreciable
reddening (mostly foreground) between us and the Cepheids in NGC~6822.
Kayser derived a reddening of E(B-V) = 0.27~mag; subsequent
determinations ranged from 0.19 to 0.42~mag (see Gallart et al. 1996).
Such a large range of reddening values suggests that a true distance
modulus based, say, on V-band data, explicitly corrected for
reddening, would immediately carry a systematic uncertainty whose 
range would amount to 0.7~mag. Indeed, most of the variance in
subsequently published distances to NGC~6822 are strongly correlated
with the adopted and/or derived reddenings.

More recent applications of the Cepheids in determining the distance
to NGC~6822, following Kayser's (1967) study, include the
multi-wavelength BVRI CCD study of six Cepheids by Gallart et
al. (1996), and the near-infrared (H-band) study of 9 Cepheids by
McAlary et al. (1983). More recently Gieren et al. (2006) obtained J
and K photometry for 56 Cepheids in NGC~6822 based on a massive
monitoring program undertaken by Pietrzynski et al. (2004) which
resulted in the cataloging of 116 variables with periods ranging from
17 to 124 days.

\section{NGC~6822: A Factor of 10 Beyond the LMC}

The SAGE observations upon which the calibration of the mid-IR PL
relations were derived from 43~seconds of integration time (for each
of two epochs), resulted in typical errors of only $\pm$0.05~mag for
stars with periods around 30~days. This would suggest that for
Cepheids a factor of ten times further away one could measure a
similar sample of long-period Cepheids to the same signal-to-noise
ratio in $43 \times 10^2$~sec, or approximately one hour.

The very nearby, Local Group, dwarf galaxy NGC~6822, is estimated to
be at a distance of 500~kpc (that is, 10 times further away than the
LMC).  There are archival sets of mid-infrared (IRAC) observations of
this galaxy consisting of 7~kilosecond mosaics. While the typical
integration time per pixel amounts to only 240~sec, this is still a
factor of about 6$\times$ longer than the LMC data used for our
calibration. These images were taken by the SINGS team for other
purposes, but they are publicly available from NED\footnote{
http://nedwww.ipac.caltech.edu/cgi-bin/nph-imgdata?objname=n6822} and
other NASA archives.  The data we used were from the SINGS fifth (and
final) data delivery.  As described in the Fifth Data Delivery
document\footnote{http://data.spitzer.caltech.edu/popular/sings/20070410\_enhanced\_v1/Documents/sings\_fifth\_delivery\_v2.pdf},
the SINGS mosaics were created from basic calibrated data (BCD) images
made by Version 14 of the SSC IRAC pipeline.  The mosaics were
drizzled to a scale of 0.75 arcsec/pixel, and use the same surface
brightness units as the original BCDs.

The IRAC images of NGC~6822 are all highly resolved into individual
stars and our first task was to identify and recover the known
Cepheids in this galaxy.  Fortunately Pietrzynski et al. (2004)
provide high-precision coordinates for all of the Cepheids discovered
by them and these naturally included the objects previously identified
by Hubble (1924) and Kayser (1967). The IRAC images have good
astrometric solutions included with them so it was a simple matter to
make preliminary identifications based solely on positional
coincidences.

At fixed resolution (2~arcsec FWHM at the shortest IRAC wavelengths)
moving a factor of ten further away (as compared to the LMC) does
raise the question as to the effects of crowding and confusion on the
photometry of NGC~6822 stars in general, and our Cepheids in
particular. It is expected that simply by going from the optical to
the mid-infrared some obvious nearby sources of crowding (blue stars)
will disappear as others (intrinsically very red stars) rise up to
take their place.  OB stars that are part of the population coeval and
co-located with the Cepheids are so blue that their effect on the
Cepheid photometry should quickly become negligible in the IRAC data.
Red supergiants, which would become even more luminous in comparison
to the Cepheids as we move from the optical to the mid-IR, are
sufficiently rare that they should not, a priori, pose much more of a
problem to these datasets in the mid-IR than they do in the
optical. The main source of confusion/crowding must then come
from asymptotic giant branch (AGB) and extended-AGB stars. These
objects are moderately luminous, very red and expected to be widely
distributed over the entire face of any galaxy having an
intermediate-aged population of stars.  Extended AGB stars are so red
that optical images are of limited use in predicting their possible
influence on a case by case basis.  In this regard we are entering
uncertain territory.

\vfill\eject 

The crowding of our sample can however be evaluated
empirically. Visual inspection of each of the Cepheids and their
immediate surroundings led to the following conclusions: nine out of
the twenty-one longest-period (P $>$ 10 days) Cepheids (NGC~6822:[K67]
V02, V05, V06, V09, V12, V14, V19, V20 and V21) in our images of
NGC~6822 were free and clear of any obvious contamination. Eight
additional objects (V01, V03, V07, V08, V10, V15, V16 and V18) have
faint nearby companions, and their photometry (especially V18 which
has been dropped from the sample) may be compromised. Four Cepheids
(V04, V11, V13 and V17) were totally crowded out at this resolution,
and the observations of V19 appears to have been affected by a bad
pixel. Depending, of course, on the color of the offending
companion(s) the degree of contamination will also be a function of
wavelength, with the reddest stars contaminating the
longest-wavelength bands the most; V07 is an example of this, where
its 8.0$\mu$m data are clearly contaminated well beyond what is seen
at shorter wavelengths. All observations that were obviously
compromised by visible companions have been dropped from further
consideration.

For these short exposures, signal-to-noise considerations also reduced
the final sample of data points available to us. The shorter-period
(fainter) Cepheids progressively failed to be detected as we moved
from the shortest to longest wavelengths. Most of the Cepheids were
either undetected or confused in the 8.0$\mu$m band, whose limiting
flux is about a factor of ten brighter than the 3.6$\mu$m limit. Only
V01, V02, V03 and V06 came through relatively unscathed at this longest
wavelength, and even then it is not clear that the longer-period star
is not contaminated at 8.0$\mu$m. Since we use only Cepheids with
periods less than 60~days or so V01 does not impact our solutions, but
then again it leaves us with only one detection in this band; and so
we focus all further discussion exclusively on the three shorter
bandpasses.

The final sample of Cepheids and their total magnitudes are given in
Table 1.  SF fitting photometry was done using DAOPHOT (Stetson
1987). Aperture corrections, derived from Table 5.7 of the IRAC Data
Handbook, were applied. The correction factors are considerable: for a
2~arcsec fitting radius they are 1.213, 1.234, 1.379 and 1.584 at 3.6,
4.5, 6.8 and 8.0$\mu$m, respectively. Finally, flux densities were
converted to the IRAC magnitude system using the zero-points of Reach
et al.  (2005) which correspond exactly to the SAGE zero points used
for the LMC calibration.

\vfill\eject

\section{IRAC Mid-Infrared Period-Luminosity Relations}

Sixteen Cepheids in NGC~6822 have good 3.6$\mu$m photometry. The
numbers monotonically decline to 14, 10 and 4 Cepheids at 4.5, 6.8 and
8.0$\mu$m, respectively. The Period-Luminosity relations based on the
data in Table 1 are shown in Figure 1. The dashed lines are
least-squares fits to the zero-point calibrating relations based on
LMC Cepheids as given in Madore et al. (2009). The flanking solid
lines are set at $\pm$0.20~mag and appear to encompass the majority of
data points. Modulo the slightly increased scatter of the NGC~6822
data as compared to the LMC ($\pm$0.14~mag) data, the calibrating
relations appear to be a good fit to the data. There is no evidence
here for any significant difference between the LMC PL slope and that
appropriate to the NGC~6822 data. 

The apparent moduli derived from fitting the LMC relations to the
NGC~6822 data are given at the bottom of Table 3. These three
independently determined moduli agree to within $\pm$0.05~mag of each
other. And, given their very low sensitivity to extinction, they
should already be a very close approximation to the true modulus for
NGC~6822.

\section{Multi-Wavelength Solutions}

Table 2 also contains a selection of additional apparent distance
moduli based on slightly different (usually larger) Cepheid samples
measured at optical through near-infrarded wavelengths. These moduli
are derived from the originally published Cepheid data, consistently
differenced against the Key Project (VI: Freedman et al. 2001, BR:
Madore \& Freedman 1991) and Persson et al. (2004) Period-Luminosity
relations at BVRI and JHK, respectively.

The moduli in Table 2 are plotted as a function of inverse wavelength
in Figure 2. Larger apparent distance (upward in the plot) as a
function of bluer wavelengths (to the right) are interpreted here as
being exclusively due to (wavelength-dependent) line-of-sight
extinction. A weighted fit to a standard Galactic entinction curve
(Cardelli et al. 1989) is shown as a solid line, flanked by one-sigma
error lines. The slope of the fit corresponds to a color excess of
E(B-V) = 0.26~mag, while the y intercept gives a true modulus of
23.49$\pm$0.03~mag. This corresponds to a metric distance of
500$\pm$6~kpc.

There is a slight discrepancy however. The two data points at J and K
(plotted as open circles) are shown but not included in the
solution. Based on near-infrared observations of over 50 Cepheids in
NGC~6822 (Gieren et al. 2006) these two data points have exceedingly
high statistical weight. But they are both systematically at variance
with the IRAC data. We do not currently have an explanation for the
difference. Gieren et al. mention applying corrections to their data
to get them onto the Persson et al. (2004) system, but those
corrections, as quoted, are small. The only other near-infrared data
point that might be used to adjudicate is the H-band measurement of
McAlary et al. (1983). This is early-epoch aperture photometry data
and of relatively low weight, but in fact it agrees to within one
sigma with both solutions. And so it is of little help here. The best
that we can do at this point is to note the problem and quantify it by
stating that ignoring the IRAC data and in deference to the JK
photometry leads to a much higher value of the reddening (E(B-V) =
0.30~mag) and consequently a lower value for the true distance
modulus, $\mu_o$ = 23.32$\pm$0.04~mag.

\section{Conclusions}

We have demonstrated that Cepheids can be detected and easily measured
by Spitzer at mid-infrared wavelengths out to at least a distance of
500~kpc with almost trivial (4~min) exposure times. Crowding (from
AGB stars primarily) is starting to become noticeable at this
distance, but sufficient numbers of Cepheids are free from crowding
that it was possible to extract uncontaminated magnitudes for over a
dozen Cepheids at 3.5$\mu$m and a monotonically decreasing number at
longer wavelengths, due to the additive affects of waning sensitivity,
slowly decreasing spatial resolution, and clearly increasing crowding and
confusion.

The four independent mid-infrared IRAC distance moduli agree to within
five percent and when combined with optical (BVRI) data they give a
true distance modulus of 23.49$\pm$0.03~mag (500~kpc) and a reddening
of E(B-V) = 0.26~mag. A discepancy (at the 0.15~mag level) with
previously published near-infrared (JK) data is noted, but no
explanation of the systematic offset is obvious.

\vfill\eject
\noindent
\centerline{\bf References \rm}
\vskip 0.1cm
\vskip 0.1cm

\par\noindent
Cardelli, J.A., Clayton, G.C.,  \& Mathis, J.S. 1989, \apj, 345, 245

\par\noindent
Freedman, W.~L., {\it et al.} 2001, \apj, 553, 47

\par\noindent Freedman, W.~L., Madore, B.F., Rigby, J., Persson, S.E.,
\& Sturch, L. 2008, \apj, 679, 71

\par\noindent
Gallart, C., Aparicio, A., \& Vilchez, J.M. 1996, \aj, 112, 1928

\par\noindent
Gieren W., et al. 2006, \apj, 647, 1056

\par\noindent
Hubble, E.P., 1925, \apj, 62, 409

\par\noindent
Kayser, S.E., 1967, \aj, 72, 134

\par\noindent
Madore, B.~F., \& Freedman, W.~L. 1991, \pasp, 103, 933

\par\noindent
Madore, B.~F., \& Freedman, W.~L. 2005, \apj, 630, 1054

\par\noindent Madore, B.F., Freedman, W.~L., Rigby, J., Persson, S.E.,
\& Sturch, L. 2009, \apj, (submitted)

\par\noindent
McAlary, C.W., Madore, B.~F., McGonegal, R., \& McLaren, R.A. 1983, \apj, 273, 539

\par\noindent
Meixner, M., {\it et al.} 2006, \aj, 132, 2268

\par\noindent
Pietrzynski, G., et al. 2004, \aj, 128, 2815

\par\noindent
Persson, S.E., et al. 2004, \aj, 128, 2239

\par\noindent
Reach, W.~T., et al. 2005, \pasp, 117, 978

\par\noindent
Rieke, G., \& Lebovsky, M. 1985, \apj, 288, 618

\par\noindent
Stetson, P. 1987, \pasp, 99, 191

\par\noindent

\vskip 0.75cm

\vfill\eject

\includegraphics [width=12cm, angle=270] {fig1.ps}

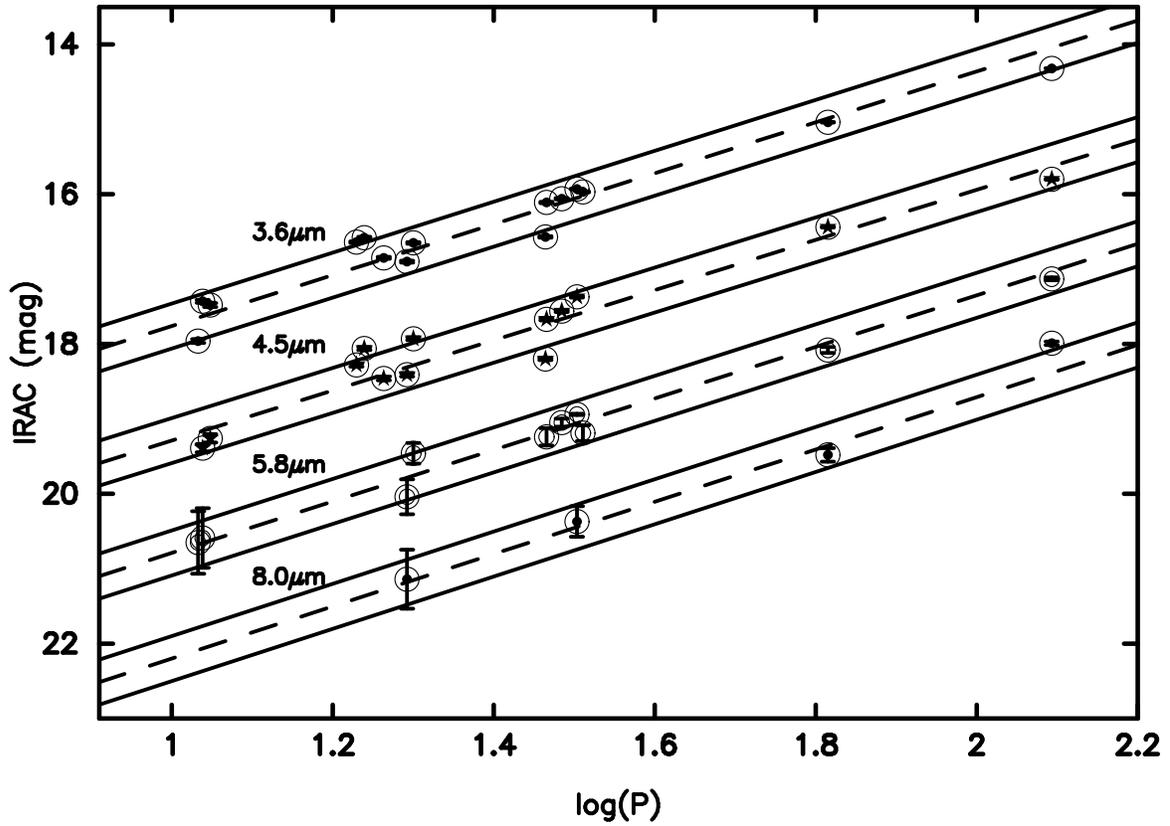
\par\noindent \bf Fig.~1 \rm -- Mid-Infrared Period-Luminosity
realtions for Cepheids in NGC~6822. Dashed lines are fits to the
fiducial relations of Madore et al. (2009) based on LMC data and an
adopted true distance to the LMC of 18.50~mag. The flanking solid
lines are offset from the fit by $\pm$0.20~mag and appear to represent
the full width of each of the relations.

\includegraphics [width=12cm, angle=270] {fig2.ps}
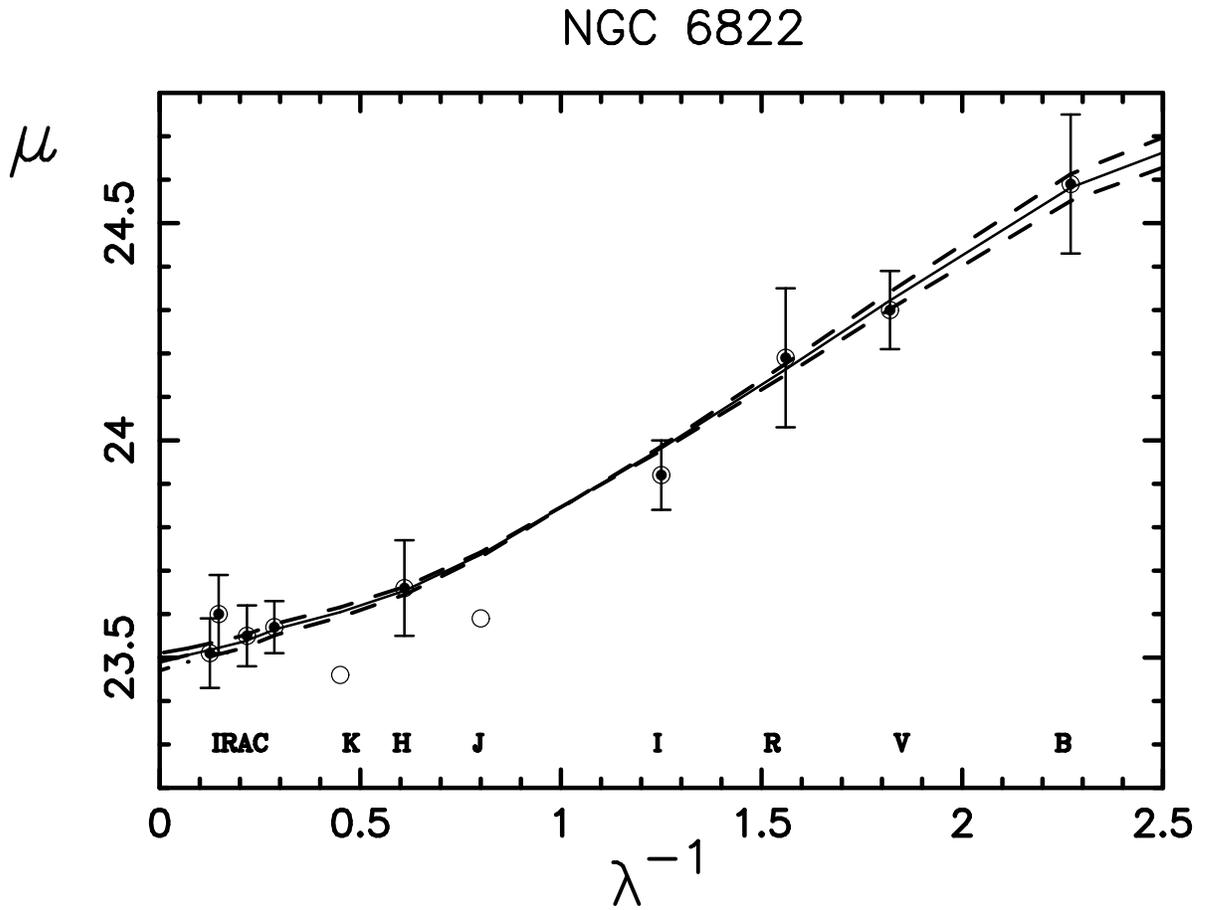
\par\noindent \bf Fig.~2 \rm -- A multi-wavelength fit of a standard
Galactic extinction curve to the apparent distance moduli to NGC~6822
derived from Cepheids. The two open circles represent J and K moduli
that are clearly at variance with the rest of the data; they are
plotted but not included in the over-all fit. The fit gives a
reddening of E(B-V) = 0.26~mag and a true distance modulus of $\mu_o$
= 23.49$\pm$0.03~mag.

\noindent

\begin{footnotesize}
\begin{deluxetable}{lccccc}
\tablecolumns{6}
\tablewidth{6.5truein}
\tablecaption{Mid-Infrared (IRAC) Magnitudes for Cepheids in NGC~6822}
\tablehead{
\colhead{Cepheid}  & \colhead{~~~~~log(P)~~~~~}  & \colhead{3.6$\mu$m}    & \colhead{4.5$\
mu$m}  & \colhead{5.8$\mu$m } & \colhead{8.0$\mu$m } 
\\ \colhead{}    & \colhead{(days)}    & \colhead{(mag)} & \colhead{(mag)}   & \colhead{(m
ag)} &\colhead{(mag)} \cr
}

\startdata
NGC~6822:[K67] V01   & 2.093 & 14.32 & 14.30 & 14.13 & 13.49 \\
      & & 0.003 & 0.006 & 0.019 & 0.024 \\
NGC~6822:[K67] V02   & 1.815 & 15.04 & 14.94 & 15.08 & 14.98 \\
      & & 0.004 & 0.005 & 0.040 & 0.090   \\
NGC~6822:[K67] V03   & 1.574 & 15.44 & 15.51 & 15.17 & 15.83 \\
      & & 0.006 & 0.009 & 0.044 & 0.182 \\
NGC~6822:[K67] V05   & 1.510 & 15.97 & . . . & 16.19 & . . . \\ 
      & & 0.006 & . . . & 0.105 & . . .  \\
NGC~6822:[K67] V06   & 1.503 & 15.93 & 15.87 & 15.94 & 15.87 \\
      & & 0.006 & 0.010 & 0.008 & 0.204 \\
NGC~6822:[K67] V07   & 1.484 & 16.06 & 16.06 & . . . & . . . \\ 
      & & 0.009 & 0.014 & . . .  & . . . \\
NGC~6822:[K67] V08   & 1.466 & 16.11 & 16.17 & 16.24 & . . . \\ 
      & & 0.009 & 0.015 & 0.116 & . . . \\
NGC~6822:[K67] V09   & 1.464 & 16.57 & 16.70 & . . . & . . . \\ 
      & & 0.010 & 0.015 & . . . & . . . \\
NGC~6822:[K67] V10   & 1.300 & 16.65 & 16.43 & 16.46 & . . . \\ 
      & & 0.011 & 0.019 & 0.138 & . . . \\
NGC~6822:[K67] V12   & 1.292 & 16.90 & 16.91 & 17.04 & . . .  \\ 
      & & 0.013 & 0.023 & 0.234 & . . . \\
NGC~6822:[K67] V14   & 1.263 & 16.85 & 16.96 & . . . & . . . \\ 
      & & 0.013 & 0.025 & . . . & . . . \\
NGC~6822:[K67] V15   & 1.239 & 16.58 & 16.56 & . . . & . . . \\ 
      & & 0.012 & 0.020 & . . . & . . . \\
NGC~6822:[K67] V16   & 1.229 & 16.64 & 16.78 & . . . & . . . \\ 
      & & 0.011 & 0.021 & . . . & . . . \\
NGC~6822:[K67] V19   & 1.048 & 17.48 & 17.76 & . . . & . . .  \\ 
      & & 0.026 & 0.056 & . . . & . . .  \\
NGC~6822:[K67] V20   & 1.038 & 17.43 & 17.89 & 17.59 & . . . \\ 
      & & 0.019 & 0.054 & 0.398 & . . . \\
NGC~6822:[K67] V21   & 1.033 & 17.96 & . . . & 17.65 & . . . \\ 
      & & 0.026& . . .  & 0.418 & . . . \\
\enddata
\end{deluxetable}
\end{footnotesize}

\begin{deluxetable}{ccc}
\tablecolumns{3}
\tablewidth{4.0truein}
\tablecaption{Apparent (Cepheid) Moduli for NGC~6822}
\tablehead{
\colhead{Bandpass}  & \colhead{Apparent Modulus}  & \colhead{Reference}
}
\startdata
 B& 24.59 (0.16) & Gallart et al. (1996)\\
 V& 24.30 (0.09) & Pietrzynski et al. (2004)\\
 R& 24.19 (0.16) & Gallart et al. (1996)\\
 I& 23.92 (0.08) & Pietrzynski et al. (2004)\\
 J& 23.59 (0.02) & Gieren et al. (2006)\\
 H& 23.66 (0.11) & McAlary et al. (1983)\\
 K& 23.46 (0.02) & Gieren et al. (2006)\\
 3.4$\mu$m& 25.57 (0.05) & this paper\\
 4.5$\mu$m& 25.55 (0.07) & this paper\\
 5.8$\mu$m& 25.60 (0.09) & this paper\\
 8.0$\mu$m& 25.51 (0.08) & this paper\\
\enddata
\end{deluxetable}

\end{document}